# Comparison between Thermoelectric and Bolometric Microwave Power Standards

Luciano Brunetti, Luca Oberto, Marco Sellone and Emil Vremera, *Member, IEEE*

*Abstract*—In the paper, a comparison is described of the microwave power standard based on thermoelectric sensors against an analogous standard based on bolometric sensors. Measurements have been carried out with the classical twin-type microcalorimeter, fitted with N-connector test ports suitable for the frequency band 0.05 – 18 GHz. An appropriate measurand definition is given for being suitable to both standard types. A system accuracy assessment is performed applying the Gaussian error propagation through the mathematical models that interpret the microcalorimeter response in each case. The results highlight advantages and weaknesses of each power standard type.

## I. INTRODUCTION

THE high frequency (HF) primary power standard is realized and maintained without alternative by means of microcalorimeters. These systems realize the standards through the measurement of the losses in dedicated mounts housing thermal detectors and provide traceability to the dc-standard at the same time [1], [2].

On our knowledge, majority of National Metrology Institutes (NMIs) still base their microcalorimeters on the bolometric detection, more or less as it was introduced in the late 1950s [3]. The most significant change concerns, perhaps, only the use of thermistors instead of resistors with positive temperature coefficient of resistance, and this both for waveguide and coaxial transmission lines [4]-[9].

However, bolometric detection does not match the industry production trend, which is oriented toward other types of power sensors, like diodes and thermocouples, and this creates some difficulties to the NMIs in replicating their HF primary power standards.

In the past, the international community did a measurement exercise, on the base of the PTB suggestions [10], by circulating power sensors based on indirect heating thermocouples among NMIs, with the aim to test the suitability of the thermoelectric detection for the coaxial microcalorimeter, as an alternative to the resistive power sensors, i.e. bolometers, no more easily available on the market at frequency beyond 18 GHz, at least [11].

The Istituto Nazionale di Ricerca Metrologica (INRIM), formerly IEN Galileo Ferraris, participated to the international comparison mentioned in [11], i.e. CIPM key comparison CCEM.RF-K10.CL - *Power in coaxial PC 3.5 mm line system*, with a coaxial microcalorimeter optimized for thermoelectric power sensors. Since then this system has been improved and its performance thoroughly studied, as reported in the wide literature of the same authors [12]-[20]. The INRIM broadband microcalorimeter, hereby mentioned, has been however designed to calibrate, without significant hardware and software changes, both classical bolometers and thermoelectric power sensors. We used this feature to perform a comparison between bolometric and thermoelectric power standards in the frequency range 0.05 -18 GHz, [21]. Hereby we report a complete comparison that highlight where and how one of the two solutions outperforms the other in realizing the HF power standard.

## II. MEASURAND DEFINITION AND MEASUREMENT SYSTEM

The measurand as subject of the comparison is the *effective efficiency $\eta_e$*, i.e. the parameter that accounts for the parasitic losses of the sensor mount and that has been well defined only for bolometers in self-balancing mode [22]. Conversely, for the thermoelectric power sensor an appropriate definition has been given for allowing the international exercise CCEM.RF-K10.CL [11].

However, independently of the sensor type, we define $\eta_e$ as ratio of the measured power $P_M$, i.e., the HF power really converted into a dc output signal by the sensor, to the total power $P_A = (P_M + P_X)$ absorbed by the sensor:

$$\eta_e = \frac{P_M}{P_M + P_X}, \qquad (1)$$

$P_X$ being the power loss in the sensor mount. This definition reduces to the typical one given in [19] for the bolometers, if we identify $P_M$ with the power measured by dc-power substitution. On other side we demonstrated in [15] that (1) matches perfectly the opportunity definition assumed in [11].

Though an improved measurement system is presently working at INRIM for power sensor calibration, we have used the same microcalorimeter that participated to the international exercise CCEM.RF-K10.CL in order to maintain a sort of data traceability to official results.

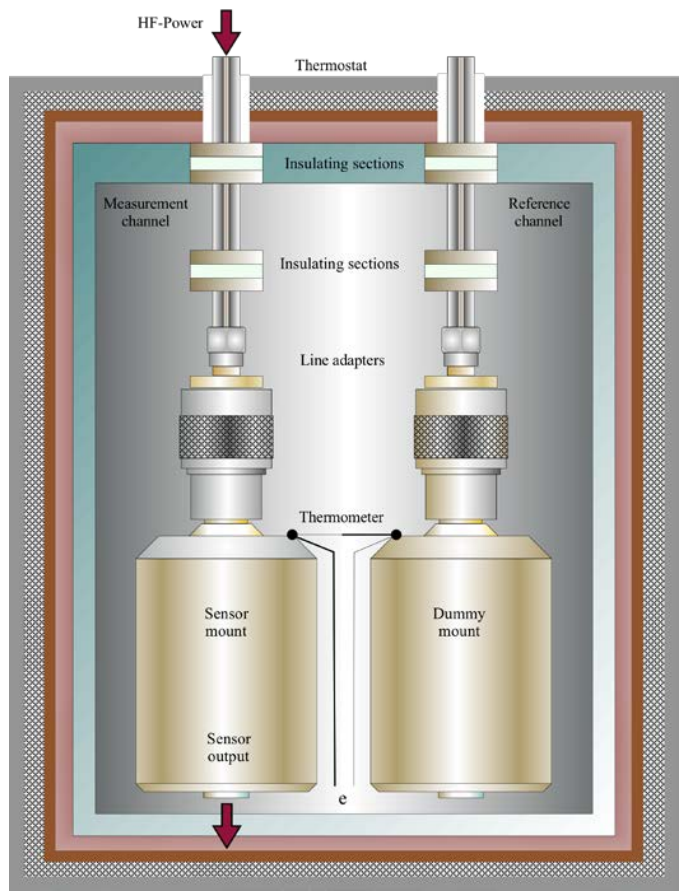

Fig. 1 Broadband microcalorimeter scheme.

The system consists in a twin-type coaxial microcalorimeter based on a triple-wall dry thermostat in which the temperature is stabilized by means of Peltier elements driven by a PID controller and acting on the intermediate wall, Fig. 1.

The microcalorimeter thermal loads are insulated from the external environment by two short adiabatic sections included in the feeding coaxial lines. The thermostat has been designed for operating inside a preconditioned room at temperature of $(23.0 \pm 0.3)$ °C and $(50 \pm 5)\%$ of relative humidity. This solution realizes a measurement chamber whose wall are maintained at about 25.0 °C with stability better of ± 0.01°C at 12 time constants, Fig 2. The time constant is that related to the time needed to reach the equilibrium temperature after the power substitution in the sensor mount. The parameter goes from 45 to 55 min, it depending on thermal capacitance of the sensor mount under calibration. The main calorimeter detector is a Cu-Constantan junction based thermometer that measures the temperature difference between the thermal loads terminating the twin insets [6]. Temperature sampling plane is at the base of the input connector of power sensors.

For performing our comparison we operated only a small hardware change on the microcalorimeter, i.e. the insertion of a PC3.5-PCN coaxial adapter to use the original 3.5 mm feeding lines.

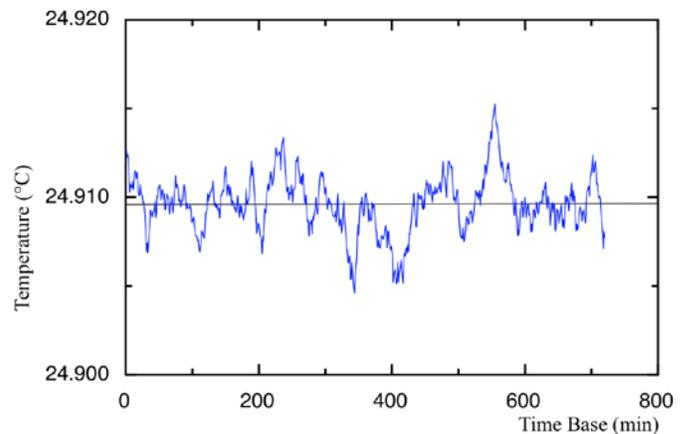

Fig. 2. Record of thermostat internal temperature T during the time requested by 4 cycles of power substitution; T can be fitted by a straight line whose angular coefficient is the temperature drift (units in $10^{-8}$ °C). Temperature mean value is $(24.910 \pm 0.001)$ °C.

### III. MICROCALORIMETER MATHEMATICAL MODELS

Though microcalorimeter has the same hardware configuration independently of the sensor type used as transfer standard, the mathematical model that links the measurand (i.e. $\eta_e$) to the temperature measurements changes according to the detection principle considered: thermoelectric or bolometric. Furthermore, power substitution follows a different way in each case and this imposes also different assumptions in the model derivation.

However, the starting point is always the principle of superimposition of the thermal effects written as follows [1]:

$$e = \alpha R(K_A P_S + K_B P_L), \quad (2)$$

where $e$ is the measure of the equilibrium temperature reached by sensor mounts after a power substitution, $\alpha$ and $R$ are dimensional conversion coefficients, $P_S$ is the total power dissipated in the sensor mount, $P_L$ is the power loss along the feeding line, $K_A$ and $K_B$ are coefficient accounting for the power rate that effectively influence the response of the electrical thermometer [15]. Expression (2) contains, however, all the influence variables that enter the determination process of $\eta_e$. By using the microcalorimeter in asymmetric mode, i.e. with only the measurement channel energized, the following mathematical models are relevant to our comparison:

#### A. Thermoelectric case

In this case measurand has proved to be [13]-[15], [17]:

$$\eta_e = \frac{e_2}{e_1 - \dfrac{e_{1SC}}{2}}, \quad (3)$$

where $e_1$, $e_2$ are, respectively, the responses of the electrical thermometer (i.e. a thermopile) to HF power and to 1 kHz reference power that is substituted on HF-feeding line. Power substitution can be done also in dc, but 1 kHz power is more appropriate for avoiding errors due to contact thermo-voltages [11], [15]. The voltage $e_{1SC}$ corrects for the microcalorimeter losses determined by means of the short circuit technique [23]. It is halved if the HF power remains the same when measuring in short circuit condition, [15]. Formula (3) holds well up to 18 GHz about because, in this range, thermoelectric power sensors exhibit a very low reflection coefficient $\Gamma_T$. If this is not the case, a further correction is necessary and term $e_{1SC}$ must be multiplied by $(1+|\Gamma_T|^2)$, [17].

*B. Bolometric case*

The effective efficiency of a thermistor mount has been deduced to be, [16]:

$$\eta_e = \frac{1-\left(\dfrac{V_{dc1}}{V_{dc2}}\right)^2}{\dfrac{e_1}{e_2}-\left(\dfrac{V_{dc1}}{V_{dc2}}\right)^2-\dfrac{M_C}{2}}, \quad (4)$$

where $e_1$, $e_2$ are the equilibrium temperatures reached by the sensor mount with and without HF power, respectively; $V_{dc1}$ and $V_{dc2}$ are the dc voltages across the bolometric element, corresponding to the dc-bias powers with and without HF; $M_C$ is the microcalorimeter correction factor to be determined by temperature measurements on the same bolometer mount under calibration after its input is short-circuited [16], [23]. According with theory, it results:

$$M_C = \frac{e_{1SC}}{e_{2SC}}-1, \quad (5)$$

where $e_{1SC}$, $e_{2SC}$ are the equilibrium temperatures reached by the bolometer mount with and without HF power, respectively. Correction factor $M_C$ must be multiplied, if it is necessary, by $(1+|\Gamma_B|^2)$, $\Gamma_B$ being the reflection coefficient of the bolometric power sensor, as for the thermoelectric case.

The asymmetry between (3) and (4) is related to the bolometric detection that has to be assisted by a dc-bias. This one eliminates the problems of sensor linearity, of course, by fixing the working point of the bolometer, but enters among the significant influence quantities.

## IV. Data Analysis

Measurement data do not enter directly (3), (4) and (5). They are elaborated by means fitting and/or averaging processes that are detailed because influencing the error budged and accuracy assessment.

The fitting process is concerning the temperature measurements, i.e. the thermopile response. This one consists in a series of increasing and decreasing exponential functions each one having a same asymptote that corresponds to a well defined equilibrium temperature of the system. The asymptote of the increasing exponential functions, $(e_1)$, is related to the excess heat produced by the HF power, while asymptote of the decreasing exponentials, $(e_2)$, relate to the equilibrium temperature of the system in presence of the reference power only (1 kHz or dc power). Figure 3 shows the thermopile response when power substitution is done in a thermoelectric sensor, whereas Fig. 4 is the analogous for the bolometer. The significant difference between them is only in the amplitude of the thermopile output, which is more than one order of magnitude higher for the bolometer. The dc bias power of the bolometer, about 30 mW, is mainly responsible of this effect.

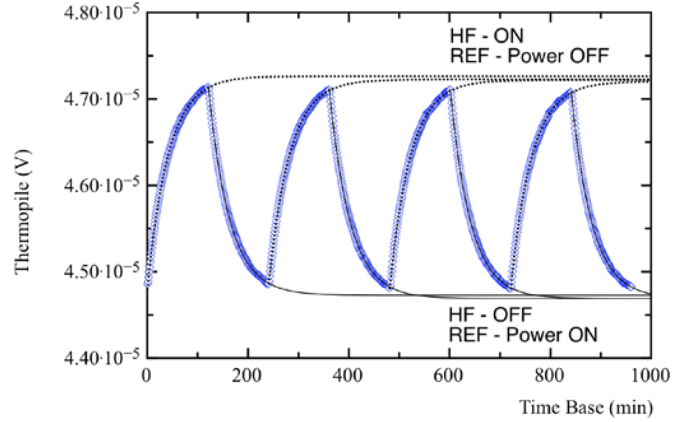

Fig. 3. Thermopile response in the thermoelectric case at 2 mW, 6 GHz; switching time 120 min.

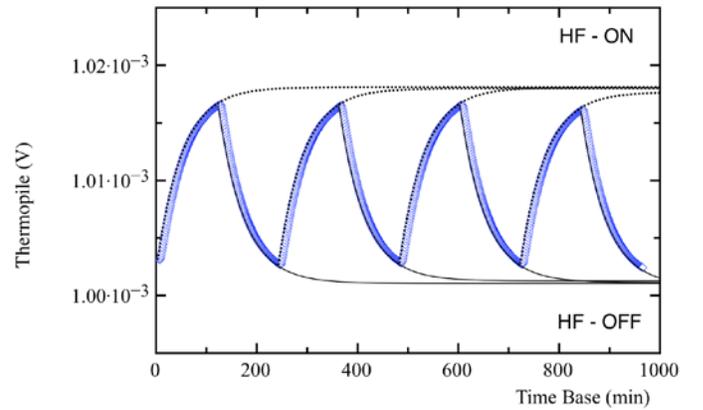

Fig. 4. Thermopile response in the bolometric case at 4 mW, 18 GHz; switching time 120 min.

Similar graphs are obtained when the microcalorimeter is operated in short circuit conditions for determining the asymptotes $e_{1SC}$ and $e_{2SC}$ that define the correction terms in (3) and (4).

The fitting method used is based on Levenberg-Marquardt algorithm, which derives directly from the mean square deviation expression [24]. Algorithm requires in input a 4-column matrix composed by the temperature measurements, i.e. the thermopile outputs, the time base values and an estimation of their measurement errors. As temperature error we considered the accuracy of the nano-voltmeter used to measure the thermopile voltage, whereas in this case, the time base has been assumed error free, because of the high precision of the sampling rate if compared to that of voltage measurements. In output we obtain the parameters of the exponential functions that compose the saw tooth signals and among them the asymptotes of our interest.

Asymptotic temperatures values $e_1$, $e_2$, $e_{1SC}$, $e_{2SC}$, obtained by fitting measurement data, are homogeneously averaged and then their mean values combined into (3), (4) and (5). The uncertainty associated to each mean value is calculated as square sum of the standard deviation of the mean and of the nano-voltmeter error. In this manner we realistically accounts both for the instrument error and for the statistic fluctuations

TABLE I
CALIBRATION LIST OF THERMOELECTRIC POWER STANDARD

| Freq. (GHz) | $\eta_e^{raw}$ | $u(\eta_e^{raw})$ | $g_T$ | $u(g_T)$ | $\eta_e$ | $U(\eta_e)$ (k = 2) |
|---|---|---|---|---|---|---|
| 0.05 | 0.9892 | 0.0004 | 1.0028 | 0.00003 | 0.9920 | 0.0009 |
| 1 | 0.9735 | 0.0004 | 1.0109 | 0.00003 | 0.9841 | 0.0009 |
| 2 | 0.9654 | 0.0004 | 1.0148 | 0.00004 | 0.9797 | 0.0007 |
| 3 | 0.9603 | 0.0004 | 1.0163 | 0.00004 | 0.9759 | 0.0006 |
| 4 | 0.9554 | 0.0005 | 1.0223 | 0.00005 | 0.9767 | 0.0006 |
| 5 | 0.9511 | 0.0005 | 1.0197 | 0.00003 | 0.9699 | 0.0007 |
| 6 | 0.9477 | 0.0003 | 1.0229 | 0.00004 | 0.9694 | 0.0008 |
| 7 | 0.9436 | 0.0005 | 1.0274 | 0.00004 | 0.9694 | 0.0010 |
| 8 | 0.9412 | 0.0003 | 1.0280 | 0.00004 | 0.9674 | 0.0009 |
| 9 | 0.9379 | 0.0004 | 1.0339 | 0.00004 | 0.9698 | 0.0007 |
| 10 | 0.9358 | 0.0003 | 1.0363 | 0.00004 | 0.9697 | 0.0007 |
| 11 | 0.9325 | 0.0004 | 1.0332 | 0.00004 | 0.9634 | 0.0010 |
| 12 | 0.9298 | 0.0003 | 1.0313 | 0.00004 | 0.9589 | 0.0009 |
| 13 | 0.9276 | 0.0004 | 1.0277 | 0.00004 | 0.9532 | 0.0009 |
| 14 | 0.9250 | 0.0003 | 1.0254 | 0.00004 | 0.9485 | 0.0015 |
| 15 | 0.9221 | 0.0005 | 1.0344 | 0.00004 | 0.9538 | 0.0010 |
| 16 | 0.9203 | 0.0005 | 1.0342 | 0.00004 | 0.9517 | 0.0009 |
| 17 | 0.9172 | 0.0004 | 1.0269 | 0.00004 | 0.9419 | 0.0008 |
| 18 | 0.9128 | 0.0003 | 1.0397 | 0.00005 | 0.9489 | 0.0010 |

TABLE II
CALIBRATION LIST OF BOLOMETRIC POWER STANDARD

| Freq. (GHz) | $\eta_e^{raw}$ | $u(\eta_e^{raw})$ | $g_B$ | $u(g_B)$ | $\eta_e$ | $U(\eta_e)$ (k = 2) |
|---|---|---|---|---|---|---|
| 0.05 | 0.9923 | 0.0016 | 1.0031 | 0.00046 | 0.9954 | 0.0042 |
| 1 | 0.9776 | 0.0016 | 1.0108 | 0.00030 | 0.9881 | 0.0038 |
| 2 | 0.9724 | 0.0024 | 1.0125 | 0.00028 | 0.9846 | 0.0053 |
| 3 | 0.9713 | 0.0033 | 1.0159 | 0.00040 | 0.9867 | 0.0074 |
| 4 | 0.9656 | 0.0029 | 1.0208 | 0.00040 | 0.9857 | 0.0067 |
| 5 | 0.9596 | 0.0018 | 1.0190 | 0.00054 | 0.9778 | 0.0046 |
| 6 | 0.9573 | 0.0041 | 1.0228 | 0.00026 | 0.9791 | 0.0090 |
| 7 | 0.9531 | 0.0031 | 1.0282 | 0.00042 | 0.9800 | 0.0073 |
| 8 | 0.9492 | 0.0017 | 1.0335 | 0.00039 | 0.9810 | 0.0042 |
| 9 | 0.9461 | 0.0029 | 1.0262 | 0.00052 | 0.9709 | 0.0062 |
| 10 | 0.9392 | 0.0028 | 1.0343 | 0.00050 | 0.9714 | 0.0067 |
| 11 | 0.9340 | 0.0022 | 1.0354 | 0.00025 | 0.9671 | 0.0052 |
| 12 | 0.9281 | 0.0014 | 1.0269 | 0.00066 | 0.9530 | 0.0042 |
| 13 | 0.9212 | 0.0022 | 1.0221 | 0.00050 | 0.9416 | 0.0056 |
| 14 | 0.9089 | 0.0028 | 1.0226 | 0.00039 | 0.9294 | 0.0064 |
| 15 | 0.8972 | 0.0014 | 1.0317 | 0.00057 | 0.9257 | 0.0040 |
| 16 | 0.8868 | 0.0015 | 1.0367 | 0.00034 | 0.9194 | 0.0037 |
| 17 | 0.8703 | 0.0017 | 1.0246 | 0.00037 | 0.8917 | 0.0041 |
| 18 | 0.8778 | 0.0017 | 1.0385 | 0.00034 | 0.9116 | 0.0041 |

of the whole system.

Data averaging is applied also to the measurements of the dc bias voltages of the bolometric element $V_{dc1}$ and $V_{dc2}$ that enter (4) and it extends appropriately to all measurement cycles, typically four. Also in this case, the uncertainty of the mean value of $V_{dc1}$ and $V_{dc2}$ is the square sum of the standard deviation of the mean and of the instrument error as given by its manufacturer.

Finally, the measurement uncertainty of the effective efficiency is determined applying the Gaussian error propagation on (3), (4) and (5), by considering however the possibility of correlations among the influence quantities and according to [25].

The described procedure gives the uncertainty with which the effective efficiency can be measured of the transfer standard and we use this parameter to qualify and to compare the thermoelectric and bolometric systems without including the contribution of the connector repeatability. Last one would require multiple connections of power sensors and further time consuming measurements, beyond the aim of this work.

## V. EXPERIMENTAL RESULTS

Power sensor calibrations have been performed at 2 mW for the thermoelectric and at 4 mW for the bolometric one, because these power levels match well the best performances of our hardware in the two cases. Tables I and II report calibration data of HF power standards, together with the uncorrected effective efficiencies $\eta_e^{raw}$ and correction factors $g_T$, $g_B$, so to highlight their effects on the final result $\eta_e$.

For the thermoelectric case $\eta_e^{raw}$ is given by the ratio $e_2/e_1$ and is related to corrected effective efficiency $\eta_e$ by [17]:

$$\eta_e = g_T \eta_e^{raw} = \left(1 - \frac{(1+|\Gamma_T|^2)}{2} \frac{e_{1SC}}{e_1}\right)^{-1} \left(\frac{e_2}{e_1}\right). \quad (6)$$

For the bolometric case, the relation between $\eta_e$ and $\eta_e^{raw}$ is [16], [19]:

$$\eta_e = g_B \eta_e^{raw} = \left(1 - \frac{\frac{1+|\Gamma_B|^2}{2}\left(\frac{e_{1SC}}{e_{2SC}}-1\right)}{\frac{e_1}{e_2}-\left(\frac{V_{dc1}}{V_{dc2}}\right)^2}\right)^{-1} \left(\frac{1-\left(\frac{V_{dc1}}{V_{dc2}}\right)^2}{\frac{e_1}{e_2}-\left(\frac{V_{dc1}}{V_{dc2}}\right)^2}\right), \quad (7)$$

where $\eta_e^{raw}$ is given by the second term in the right hand side. Both (6) and (7) are models that accounts for all the significant systematic errors respectively in thermoelectric and bolometric case [16]-[23]. Expanded uncertainty terms $U(\eta_e)$ in Tables I and II are given with coverage factor k=2.

A first analysis of the comparison can be done from Fig. 5 and 6 that report the results with error bars, even though they are not always evident because of the graph scales. In the thermoelectric case, corrected effective efficiencies $\eta_e$ exhibit much lower uncertainties than in the bolometric case (see also the numeric values in Tables I and II). Furthermore, the raw effective efficiencies $\eta_e^{raw}$ and the correction factors $g_T$, $g_B$ have same characteristic behavior.

Concerning $g_T$ and $g_B$, Fig. 6 shows also that their trend versus frequency is almost the same, thing confirmed by a calculated correlation coefficient greater than 0.95 and a sum of the squared differences less than 0.0002. We attribute the discrepancies that appear in some points to a low repeatability of the short circuit connection and to tear and wear of the microcalorimeter test port. Figure 6 suggests that the microcalorimeter maintains its performances independently of the sensor type used as load, but data series of Tables I and II clearly shows that the whole realization process of the primary power standard is more accurate if the microcalorimeter load is of thermoelectric type. In this sense, thermoelectric sensors outperform bolometers when used as microcalorimeter load.

TABLE III
DETAILS OF ERROR BUDGET AT 10 GHZ FOR THERMOELECTRIC AND BOLOMETRIC POWER STANDARDS (EXCLUDING ADIMENSIONAL REFLECTION COEFFICIENTS, QUANTITIES AND RELATED UNCERTAINTIES ARE IN VOLT)

| Influence Variable | Measured Value $y$ | Measurement Uncertainty $u(y)$ | Sensitivity coefficient $|c(y)|$ | Uncertainty Contribution $c(y)u(y)$ |
|---|---|---|---|---|
| *Thermoelectric Standard* | | | | |
| $e_1$ | 4.8090E-05 | 1.3799E-08 | 2.0973E+04 | 0.00029 |
| $e_2$ | 4.4800E-05 | 1.0563E-08 | 2.1637E+04 | 0.00023 |
| $e_{1SC}$ | 0.3749E-05 | 4.2558E-09 | 1.0488E+04 | 0.00004 |
| $\Gamma_T$ | 0.0126 | 0.0080 | 0.0010 | 0.00001 |
| $U(\eta_e)$ ; (k=2) | | | | 0.00074 |
| *Bolometric Standard* | | | | |
| $e_1$ | 1.0108E-03 | 2.9301E-07 | 7.2645E+03 | 0.00213 |
| $e_2$ | 1.0023E-03 | 2.9030E-07 | 7.3261E+03 | 0.00213 |
| $e_{1SC}$ | 1.0068E-03 | 3.2636E-07 | 3.6585E+03 | 0.00119 |
| $e_{2SC}$ | 0.9962E-03 | 2.4708E-07 | 3.6975E+03 | 0.00091 |
| $\Gamma_B$ | 0.0341 | 0.0080 | 0.0026 | 0.00002 |
| $V_{dc1}$ | 2.294811 | 0.000056 | 0.132824 | 0.00001 |
| $V_{dc2}$ | 2.461658 | 0.000059 | 0.123822 | 0.00001 |
| $U(\eta_e)$ ; (k=2) | | | | 0.00673 |

To understand why this happens, it is necessary to examine with more detail how the error budged forms at a single frequency. Table III shows uncertainty contributions given by microcalorimeter models (6), (7) in the power standard realization at 10 GHz, pointing out that any other lower or higher frequency exhibits the same behavior. Covariant terms are not reported in Table III because resulted negligible compared to main error terms related to temperature measurements $e_1$, $e_2$ during the sensor calibration phase and $e_{1SC}$, $e_{2SC}$ in the microcalorimeter calibration phase. Table III shows that only the temperature measurements give significant error contributions both in the thermoelectric case and in the bolometric one. However, combinations of measurement uncertainties and sensitivity coefficients are more favorable for the thermoelectric standard. This is an intrinsic consequence of the mathematical models used (6) and (7) which have a different number of influence variables.

Furthermore, we can infer that the power sensor mismatch has small impact on the results. The real significant systematic error source is due to the feeding line losses and the effect for their correction is evidenced in Fig. 5 that reports, on same scale, raw and corrected effective efficiencies of the two power sensor types. The effects of such losses are evaluated through the thermo-voltages $e_{1SC}$ and $e_{2SC}$ that are generally difficult to measure because very small and close to the noise floor of the system (e.g. around 30 nV in the thermoelectric power sensor case). However, despite this inconvenient, the accuracy of the microcalorimeter calibration process, i.e. the $g_B$ and $g_T$ determination, is not critically influenced.

Further assessment of these results can be only trough an international comparison in which the reference values are stated by measurements made with microcalorimeters. Any other measurement technique can return only higher measurement uncertainties that mask intrinsic properties of the model used [26].

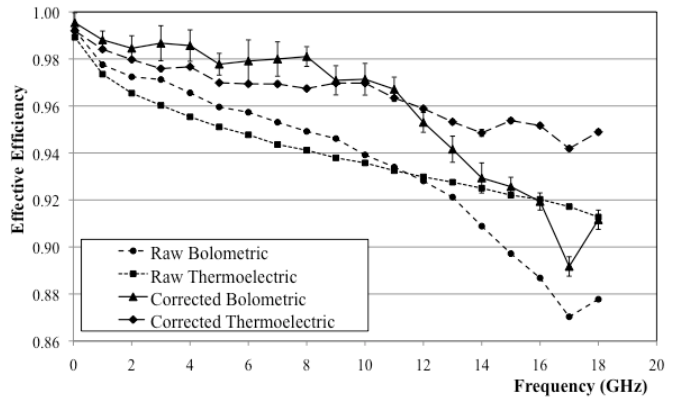

Fig. 5. Trend of effective efficiencies (raw and corrected) in the case of thermoelectric and bolometric HF power standard. Error bars are given for a coverage factor k=2.

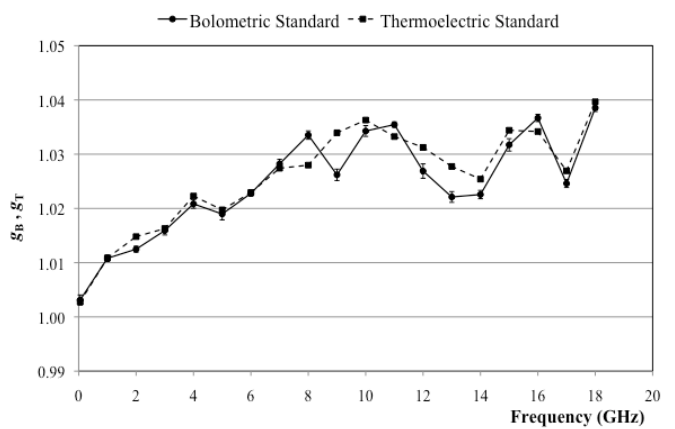

Fig. 6. Comparison of the correction factors of the microcalorimeter $g_B$ and $g_T$ for bolometric and thermoelectric configuration. Error bars are given for a coverage factor k=2.

IV. CONCLUSION

A comparison between thermoelectric and bolometric sensors made with the same microcalorimeter, revealed that the realized two primary power standards can be considered equivalent, but the thermoelectric version results with an intrinsic higher accuracy. Moreover thermoelectric standard has higher effective efficiency values in the upper part of the considered band and this because the power sensor type used, i.e. indirect heating thermocouple, does not suffers the problem of the high frequency leakage through a dc bias circuit, like for the bolometers. Bolometric sensors, that are much more sensitive to absolute temperature variations, can be considered interesting only for being independent of the power level, that is, free of linearity error. Conversely, for the thermoelectric sensor, this error must be considered and the simplest way for working around is to operate at power level possibly lower than 3 mW and of course compatible with the sensitivity of the own measurement system.

Finally, the thermoelectric detection is eligible to become the reference in the specific world of the HF primary metrology, solving in this manner also problems of the difficult procurement of bolometers on the present market.

## REFERENCES

[1] A. Fantom, *Radiofrequency µwave power measurement*, London, U. K. : Peter Peregrinus Ltd.,1990.

[2] R. Clark, "The microcalorimeter as a national microwave standard", *Proc. IEEE.*, vol. 74, No. 1, Jan. 1986, pp. 104–122.

[3] A. C. Macpherson and D. M. Kerns, "A Microwave Microcalorimeter", *Rev. Sci. Instr.*, vol. 26, no. 1, pp. 27-33, Jan. 1955.

[4] N. S. Chung, J. Sing, H. Bayer and R. Honigbaum, "Coaxial and waveguide microcalorimeters for RF and microwave power standards", *IEEE Trans. Instr. Meas.*, vol. 38, no. 2, pp. 460-464, Apr. 1989.

[5] F. R. Clague, "A method to determine the calorimetric equivalence correction for a coaxial microwave microcalorimeter ", *IEEE Trans. Instrum. Meas.*, vol. 43, no. 3, pp. 421-425, Jun. 1994.

[6] A. Akhiezer, A. Senko and V. Seredniy, "Millimeter wave power standards", *IEEE Trans. Instr. Meas.*, vol. 46, pp. 495–498, Apr. 1997.

[7] Y. Okano and T. Inoue, "Automatic Microcalorimeter System for Broadband Power Measurement in 75 GHz - 110 GHz Range", *IEEE Trans. Instr. Meas.*, vol. 58, no. 2, pp. 385–388, Apr. 2001.

[8] J.P.M. de Vreede, "Final report of the comparison CCEM.RF-K8.CL: calibration factor of thermistor mount", *Metrologia*, vol. 42, Technical Supplement, 01008, 2005.

[9] R. Judaschke and J. Rühaak, "Determination of the Correction Factor of Waveguide Microcalorimeters in the Millimeter-Wave Range", *IEEE Trans. Instr. Meas.*, vol. 58, no. 4, pp. 1104-1108, Apr. 2009.

[10] E. Vollomer, *et. al.*, "Microcalorimeter measurement of the effective efficiency of microwave power sensors comprising thermocouples", in *CPEM Conf. Dig.,* 1994, pp. 147–148.

[11] D. Janik, et. al., "Final report on CCEM Key Comparison CCEM.RF-K10.CL (GT/RF 99-2); Power in 50 Ohm coaxial lines, frequency: 50 MHz to 26 GHz – measurement techniques and results", *Metrologia*, vol. 43, Technical Supplement, 01009, 2006.

[12] L. Brunetti, Y. Shan, L. Oberto, C. W. Chua, M. Sellone and P. Terzi, "High frequency comparison with thermoelectric power sensors between INRIM and NMC", *Measurement*, vol. 45, pp. 1180-1187, Feb. 2012.

[13] L. Brunetti and E. Vremera, "A new microcalorimeter for measurements in 3.5-mm coaxial line", *IEEE Trans. Instr. Meas.*, vol. 52, No. 2, pp. 320–323, Apr. 2003.

[14] L. Brunetti, L. Oberto and E. Vremera, "Thermoelectric sensor as microcalorimeter loads", *IEEE Trans. Instr. Meas.*, vol. 56, No. 6, pp. 2220–2224, Dec. 2007.

[15] L. Brunetti, L. Oberto, M. Sellone and E. Vremera, "Comparison among coaxial microcalorimeter models", *IEEE Trans. Instr. Meas.*, vol. 58, No. 4, pp. 1141–1145, Apr. 2009.

[16] L. Brunetti and E. Vremera, "A new calibration method for coaxial microcalorimeters", *IEEE Trans. Instr. Meas.*, vol. 54, No. 2, pp. 684–685, Apr. 2005.

[17] L. Brunetti, L. Oberto, M. Sellone and E. Vremera, "Latest determination of a coaxial microcalorimeter calibration factor", *Meas. Sci. Technol.*, vol. 22, pp. 025101-6, Dec. 2011.

[18] E. Vremera, L. Brunetti, L. Oberto and M. Sellone, "Alternative procedure in realizing of the high frequency power standards with microcalorimeter and thermoelectric detectors", *Measurement*, vol. 42, pp. 269-276, Feb. 2009.

[19] L. Oberto, L. Brunetti and M. Sellone, "True-twin microcalorimeter: proof-of-concept experiment", *Electr. Lett.*, vol. 47, No. 9, pp. 550-551, Apr. 2011.

[20] E. T. Vremera, L. Brunetti, L. Oberto and M. Sellone, "Power sensor calibration by implementing true-twin microcalorimeter", *IEEE Trans. Instr. Meas.*, vol. 60, No. 7, pp. 2335-2340, Jul. 2011.

[21] L. Brunetti, L. Oberto, M. Sellone and E. Vremera, "Thermoelectric against bolometric microwave power standard", in *CPEM Conf. Dig.*, 2012.

[22] *IEEE Standard Application Guide for Bolometric Power Meters*, IEEE Standards 470, 1972.

[23] L. Brunetti and L. Oberto, "On coaxial microcalorimeter calibration", Eur. Phys. J. Appl. Phys., vol. 43, pp. 239-244, 2008.

[24] W. H. Press, B. P Flannery, S. A. Teukolsky and W.T. Vetterling, *Numerical Recipes – the Art of Scientific Computing*, Second Edition, University Press, Cambridge, 1992.

[25] BIPM, IEC, IFCC, ISO, IUPAC, IUPAP and OIML, *Guide to the Expression of the Uncertainty in Measurement, 1995,* 2nd ed.

[26] M. Sellone, L. Brunetti, L. Oberto and P. Terzi, "Realization and dissemination of high frequency power standard at INRIM", *Measurement*, vol.45, pp. 290-296, Apr. 2012.

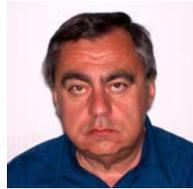

**Luciano Brunetti** was born in Asti, Italy, on September 11, 1951. He received the M.S. degree in Physics from the University of Torino, Italy, in 1977.

Since 1977 he has been working at Istituto Nazionale di Ricerca Metrologica (INRIM, formerly IEN "Galileo Ferraris"), Torino, Italy. He has been dealing both with theoretical and experimental research in the field of high frequency primary metrology. His main task has always been the realization and the dissemination of the national standard of power, impedance and attenuation in the microwave range. In the last years he has been involved in the design and characterization of millimeter and microwave devices working at cryogenic temperature; he collaborates also at the characterization of complex magnetic alloys at high frequency. Actually he is taking care of the extension of the national electrical standards in the millimeter wavelength range.

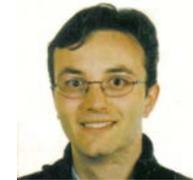

**Luca Oberto** was born in Pinerolo (Torino), Italy, on June 9, 1975. He received the M.S. degree in Physics from the University of Torino in 2003 and the Ph.D. in Metrology from the Politecnico di Torino in 2008.

From 2002 to 2003 he was with the Istituto Nazionale di Fisica Nucleare (INFN), Torino Section, working at the COMPASS experiment at CERN, Geneva, Switzerland. From 2003 he is with the Istituto Nazionale di Ricerca Metrologica (INRIM), Torino, Italy. His research interests are in the field of high frequency and THz metrology and in the realization and characterization of superconductor-insulator-superconductor mixers for astrophysical applications in the millimeter- and sub millimeter-wave domain.

Dr. Oberto is member of the Associazione Italiana Gruppo di Misure Elettriche ed Elettroniche (GMEE). He was recipient of the 2008 Conference on Precision Electromagnetic Measurements Early Career Award and of the GMEE 2010 "Carlo Offelli" Ph.D. prize.

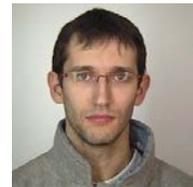

**Marco Sellone** was born in Turin, Italy, on September 25, 1979. He received the M.S. degree in Physics from the University of Torino, Italy, in 2004 and the Ph.D. degree in Metrology from the Politecnico di Torino in 2009.

Since 2005, he has been with the Istituto Nazionale di Ricerca Metrologica (INRIM), Torino, Italy. His research interests include the high-frequency metrology mainly regarding ac–dc transfer difference measurements, working on the development of a new measurement setup to extend INRIM ac–dc transfer difference measurement capabilities above 1 MHz, and high-frequency standards, particularly vector network analyzer for scattering parameter measurements.

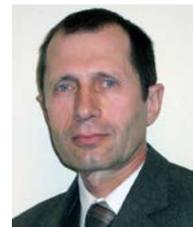

**Emil T. Vremera** (M'07) was born in Romani-Neamt, Romania, on September 11, 1953. He received the M.Sc. degree in electronics and the Ph.D. degree in electrical measurements from "Gheorghe Asachi" Technical University of Iasi, Romania, in 1977 and 1998, respectively.

He is currently with the Department of Electric Measurements, Faculty of Electrical Engineering, "Gheorghe Asachi" Technical University of Iasi, where he joined in 1984 first as an Assistant Professor and then as a Professor. He teaches electric and electronic measurements for the students in the electronic area. Since 2001, he has been developing a research activity on RF power measurements. He is also an Associated Scientist with Istituto Nazionale di Ricerca Metrologica, Turin, Italy. His main research interests concern measurement techniques of the electric and magnetic quantities, analog to digital conversion for second-order quantities, and virtual instrumentation.